\documentclass[twocolumn,showpacs,preprintnumbers,amsmath,amssymb,prl,superscriptaddress]{revtex4}
\usepackage{graphicx}

\begin{document}

\title{Scattering of charge carriers by point defects in bilayer graphene}
\author{M. I. Katsnelson}
\email{M.Katsnelson@science.ru.nl} \affiliation{Institute for
Molecules and Materials, Radboud University Nijmegen, 6525 ED
Nijmegen, The Netherlands}


\begin{abstract}
Theory of scattering of massive chiral fermions in bilayer
graphene by radial symmetric potential is developed. It is shown
that in the case when the electron wavelength is much larger than
the radius of the potential the scattering cross-section is
proportional to the electron wavelength. This leads to the
mobility independent on the electron concentration. In contrast
with the case of single-layer, neutral and charged defects are, in
general, equally relevant for the resistivity of the bilayer
graphene.
\end{abstract}

\maketitle

\affiliation{Institute for Molecules and Materials, Radboud
University Nijmegen, 6525 ED Nijmegen, The Netherlands}

Bilayer graphene, that is a two-dimensional allotrope of carbon
formed by two graphite atomic sheets \cite{natphys}, is a subject
of hot interest now
\cite{falko,klein,castro,andobi,gap,land1,land2,peeters,falkoSSC,wl1,wl2}
motivated by anomalous character of the quantum Hall effect
\cite{natphys,falko} and electron transmission through potential
barriers \cite{klein} due to electron chirality and the Berry
phase $2\pi $, possible use of the bilayer graphene as a
tunable-gap semiconductor \cite{gap} and its other unusual
physical properties (for review, see Ref.\onlinecite{falkoSSC}).
At the same time, it is less studied than the single-layer
graphene \cite{reviewGK,reviewktsn}. In particular, almost nothing
is known about mechanisms of scattering determining the electron
transport in the bilayer graphene. Here we consider this problem
theoretically. It will be shown that for any kind of point defects
with small enough concentration their contribution in the
resistivity is inversely proportional to the charge carrier
concentration resulting in the concentration-independent electron
mobility. In a framework of perturbation theory, this result has
been obtained earlier in Ref.\onlinecite{andobi} (see their
Eq.(52)); we have generalized it on a case of a strong impurity
potential. This situation is essentially different from the
single-layer case when the scattering by Coulomb potential of
charge impurities leads to the concentration-independent mobility
whereas the short-range scattering centers are almost irrelevant
\cite{nomura,ando,sarma,SSC}.

The bilayer graphene in a simplest approximation can be considered
as a zero-gap semiconductor with parabolic touching of the
electron and hole bands described by the single-particle
Hamiltonian \cite{natphys,falko,falkoSSC}
\begin{equation}
H=\left(
\begin{array}{cc}
0 & -\left( p_x-ip_y\right) ^2/2m \\
-\left( p_x+ip_y\right) ^2/2m & 0
\end{array}
\right)   \label{bilayer}
\end{equation}
where $p_i=-i\hbar \partial /\partial x_i$ are electron momenta
operators and $m\simeq 0.054m_e$ is the effective mass, $m_e$
being the free-electron mass. This description is accurate at the
energy scale larger than few meV, otherwise a more complicated
picture including trigonal warping takes place; we will restrict
ourselves only by the case of not too small doping when the
approximate Hamiltonian (\ref{bilayer}) works. Two components of
the wave function are originated from crystallographic structure
of graphite sheets with two carbon atoms in the sheet per
elementary cell. There are two touching points per Brillouin zone,
$K$ and $K^{\prime }$. For smooth enough external potential, no
Umklapp processes between these points are allowed and thus they
can be considered independently.

The Fourier component of the impurity potential with dimensionless charge $Z$
at small enough wave vector equals
\begin{equation}
V\left( q\right) =\frac{2\pi Ze^2}{\epsilon \left( q+\kappa \right) },
\label{kappa}
\end{equation}
where $\kappa =2\pi e^2N\left( E_F\right) /\epsilon $ is the
inverse screening radius, $\epsilon \simeq 2.5$ is the dielectric
constant due to quartz substrate, and $N\left( E_F\right) $ is the
density of states at
the Fermi energy $E_F$ \cite{nomura,ando}. In the model (\ref{bilayer}) $%
\kappa =4me^2/\hbar ^2\epsilon $ where we take into account
contributions from two spin projections and two valleys. Due to
the smallness of the effective mass the screening radius is 4.5
times larger than the nearest-neighbor interatomic distances which
makes the single valley approximation accurate enough. At the same
time, for any reasonable doping the Fermi wave vector $k_F\ll
\kappa $ so one can assume that the electron wavelength is much
larger than the scattering potential radius.

Let us consider the case of small concentration of point defects
(to be specific, we will call them impurities) with the
concentration $n_{imp}$ and the angle-dependent scattering
cross-section $\sigma \left( \phi \right) .$ Then the defect
contribution to the resistivity $\rho $ reads~\cite{ziman,shon}
\begin{eqnarray}
\rho  &=&\frac 2{e^2v_F^2N\left( E_F\right) }\frac 1{\tau \left( k_F\right)
},  \nonumber \\
\frac 1{\tau \left( k_F\right) } &=&n_{imp}v_F\int\limits_0^{2\pi }d\phi
\frac{d\sigma \left( \phi \right) }{d\phi }\left( 1-\cos \phi \right)
\label{resist}
\end{eqnarray}
where $v_F=\hbar k_F/m$ is the Fermi velocity, $\tau $ is the
mean-free-path time. Note that the product $v_FN\left( E_F\right)
$ is proportional to $k_F= \sqrt{\pi n}$ ($n$ is the electron
concentration) for both single-layer and bilayer graphene, as well
as for conventional two-dimensional electron gas and thus any
essential difference in their transport properties can be related
only to the behavior of the scattering cross-section.

The expression (\ref{resist}) is derived from the standard
Boltzmann equation and does not take into account localization (or
antilocalization) corrections which can change the results
drastically in the regime of small doping when the resistivity is
of order of $h/e^2$ \cite{altshuler,mirlin,efetov,altland}
(recently the problem of weak localization has been considered
also for the bilayer graphene \cite{wl2}). We will restrict
ourselves only by the case $\rho \gg h/e^2$; formal derivation of
the Boltzmann equation for graphene in this regime will be
published elsewhere \cite{auslender}.

To determine the scattering cross section one has to solve the
two-dimensional Schr\"odinger equation with the Hamiltonian
(\ref{bilayer}) plus impurity potential $V\left( r\right) $ which,
after simple manipulations (cf. Ref.\onlinecite{peeters}) can be
written in the form
\begin{eqnarray}
\left( \frac d{dr}- \frac{l+1}r\right) \left( \frac d{dr}-\frac
lr\right)
g_l &=& \left( k^2-\frac{2mV }{\hbar ^2}%
\right) f_l ,  \nonumber \\
\left( \frac d{dr} + \frac{l+1}r\right) \left( \frac
d{dr}+\frac{l+2}r\right)
f_l &=& \left( k^2-\frac{2mV }{\hbar ^2}%
\right) g_l . \nonumber \\
\label{radial}
\end{eqnarray}
where $l=0,\pm 1,...$ is the angular-momentum quantum number,
$g_l\left( r\right) e^{il\phi }$ and $f_l\left( r\right)
e^{i(l+2)\phi }$ are components of the pseudospinor wave function,
$r$ and $\phi $ are polar coordinates; to be specific we will
consider the case of electrons $E=\hbar ^2k^2/2m>0$.

Modifying a standard scattering theory~\cite{newton} for the two-dimensional
case one should try the solutions of Eq.(\ref{radial}) \ outside the region
of action of the potential in the form
\begin{eqnarray}
g_l\left( r\right)  &=&A\left[ J_l\left( kr\right) +t_lH_l^{(1)}\left(
kr\right) +c_lK_l\left( kr\right) \right] ,  \nonumber \\
f_l\left( r\right)  &=&A\left[ J_{l+2}\left( kr\right)
+t_lH_{l+2}^{(1)}\left( kr\right) +c_lK_{l+2}\left(
kr\right)\right], \nonumber \\
\label{bessel}
\end{eqnarray}
where the terms proportional to Bessel (Hankel) functions describe
incident (scattering) waves; the terms proportional to the
Macdonald functions are analogous to the exponentially decaying
solutions in the case of potential barrier \cite{klein}. To
calculate the scattering cross
section one has to find the current operator $\mathbf{j=}\frac 1\hbar \frac{%
\delta H}{\delta \mathbf{k}}$ and its normal component $j_n=j_x\cos \phi
+j_y\sin \phi .$ The result reads:
\begin{equation}
j_n=-\frac{\hbar k}m\left(
\begin{array}{cc}
0 & e^{-2i\phi } \\
e^{2i\phi } & 0
\end{array}
\right) .  \label{current}
\end{equation}

The Bessel and Hankel functions in Eq.(\ref{bessel}) correspond to
the expansion of the incident plane wave and scattered radial
wave, respectively. Calculating the average value of the current
operator (\ref{current}) over the scattered wave we find for the
cross section
\begin{equation}
\frac{d\sigma \left( \phi \right) }{d\phi }=\frac 2{\pi k}\left|
\sum\limits_{l=-\infty }^\infty t_le^{il\phi }\right| ^2,  \label{crosssect}
\end{equation}
which is formally the same expression as for the case of single-layer
graphene \cite{SSC}.

The Schr\"odinger equation (\ref{radial}) has as important symmetry with
respect to replacement $f\longleftrightarrow g,l\longleftrightarrow -l-2$
which means $t_l=t_{-l-2}.$ This is the consequence of chiral properties of
electrons with the Berry phase $2\pi $; a similar identity for the
single-layer case with the Berry phase $\pi $ reads \cite{SSC} $%
t_l=t_{-l-1}$. Thus, Eq.(\ref{crosssect}) can be rewritten in the
form
\begin{equation}
\frac{d\sigma \left( \phi \right) }{d\phi }=\frac 2{\pi k}\left|
t_{-1}+2\sum\limits_{l=0}^\infty t_l\cos \left[ \left( l+1\right)
\phi \right] \right| ^2.  \label{crosssect1}
\end{equation}

To understand the behavior of the scattering parameters $t_l\left(
k\right) $ in the interesting limit $k\rightarrow 0$ one can
consider the simplest case of the potential $V\left( r\right)
=V_0$ at $r<a$ and $V\left( r\right) =0$ at $r>a.$ Strictly
speaking, a sharp jump of the potential with atomic scale is
beyond applicability of our approach since it will induce Umklapp
processes between the valleys. We assume that the boundary is
smooth enough in comparison with the interatomic distance but much
thinner than the electron wavelength (cf. Ref.\onlinecite{klein}).
The solution outside the potential well has the form (\ref
{bessel}), with $A=1$ and the solution for $r<a$ regular at $r=0$
can be tried as
\begin{eqnarray}
g_l\left( r\right)  &=&\alpha _lJ_l\left( qr\right) +\beta _lI_l\left(
qr\right) ,  \nonumber  \label{bessel11} \\
f_l\left( r\right)  &=&\sigma \left[\alpha _l J_{l+2}\left(
qr\right) +\beta _lI_{l+2}\left( qr\right) \right] ,
\label{bessel11}
\end{eqnarray}
where $\sigma =sign\left( E-V_0\right)$  and $q=\sqrt{2m\left| E-V_0\right| }%
/\hbar $ is the wave vector inside the well. Using boundary
conditions of continuity of the wave functions and their first
derivatives at $r=a$ one
can find the scattering parameters $t_l$ as well as $c_l,\alpha _l$ and $%
\beta _l$ (cf. the case of one-dimensional potential
\cite{klein}).

For the case $l=-1$ taking into account identities $K_1\left(
z\right) =K_{-1}\left( z\right) ,I_1\left( z\right) =I_{-1}\left(
z\right) ,J_1\left( z\right) =-J_{-1}\left( z\right) ,$and
$H_1^{(1)}\left( z\right) =-H_{-1}^{(1)}\left( z\right) $ one can
prove immediately that $c_{-1}=0$ and $t_{-1}\propto \left(
ka\right) ^2$ at $ka\rightarrow 0$ so as we will see this
contribution in the scattering cross section is negligible.
Using asymptotic of the Macdonald and Hankel functions for $%
l>2,z\rightarrow 0$%
\begin{eqnarray}
K_l\left( z\right)  &\simeq &\frac 12\left( \frac 2z\right)
^l\left( l-1\right) !-\frac 12\left( \frac 2z\right) ^{l-2}\left(
l-2\right) !,
\nonumber \\
H^{(1)}_l\left( z\right)  &\simeq &-\frac i{\pi }\left( \frac
2z\right) ^l\left( l-1\right) !-\frac i{\pi }\left( \frac
2z\right)
^{l-2}\left( l-2\right) !, \nonumber \\
\end{eqnarray}
one can prove that for $l\geq 1$ and $ka\rightarrow 0$ both $t_l$ and $c_l$
are, at least, of order of $\left( ka\right) ^{2l}$or smaller and thus only $%
s$-channel $\left( l=0\right) $ contributes in the scattering
cross section so that Eq.(\ref{crosssect1}) can be rewritten as
\begin{equation}
\frac{d\sigma \left( \phi \right) }{d\phi }=\frac 8{\pi k}\left|
t_0\left( k\right) \right| ^2\cos ^2\phi .  \label{crosssect2}
\end{equation}
For the single-layer graphene, $\sigma \left( \phi \right) \propto
\cos ^2{\phi/2} $ and the back-scattering is forbidden. On the
contrary, for the case of bilayer there is a strong suppression of
the scattering at $\phi \simeq \pi/2$.

For the case $l=0$ the wavefunctions outside the well (\ref{bessel}) has the
asymptotic form
\begin{eqnarray}
g_l\left( r\right)  &=&1+t_0+\tau _0\left( \ln \frac{kr}2+\gamma \right) +%
\mathcal{O}\left( \left( kr\right) ^2\ln kr\right) ,
\nonumber\\
f_l\left( r\right)  &=&-\frac{2i}\pi t_0-\tau _0\left( \frac
2{\left( kr\right) ^2}-\frac 12\right) +\mathcal{O}\left( \left(
kr\right) ^2\ln kr\right) \nonumber \\
\end{eqnarray}
where $\gamma \simeq 0.577...$ is the Euler constant, $\tau
_0=\frac{2i}\pi t_0-c_0.$ Using this we find that $t_0\left(
k\right) $ tends to a finite complex number $\left( \left|
t_0\left( k\right) \right| ^2\leq 1\right) $ at $k\rightarrow 0$.
Substituting this into Eqs.(\ref{crosssect2}) and (\ref {resist})
one can find an estimation for the resistivity  $\rho \simeq
\left( h/4e^2\right) n_{imp}/n.$ It seems to be in a qualitative
agreement with the dependence of the resistance of bilayer
graphene on the gate voltage measured in Ref.\onlinecite{wl1}. The
same dependence of the resistivity on the charge carrier
concentration takes place for the single-layer graphene with
Coulomb scattering centers \cite{nomura} whereas the point defects
with short-range potential give much smaller resistivity
\cite{SSC} of order of $\rho \simeq \left( h/4e^2\right)
n_{imp}a^2.$ For the case of bilayer, on the contrary, there is,
in general, no essential difference between charge impurities and
neutral point defects such as, say, atomic-scale roughness of the
substrate.

It is interesting to mention that the scattering by the
short-range potential in the case of bilayer graphene is more
efficient than not only in the case of the single-layer graphene
but also for the conventional nonrelativistic two-dimensional
electron gas where $t_0\left( k\right) \propto 1/\ln \left(
ka\right) $ at $ka\rightarrow 0$ and thus \cite
{adhi,SSC,misprint}
\begin{equation}
\rho \simeq \frac h{4e^2}\frac{n_{imp}}{n\ln ^2\left( k_Fa\right) }
\label{resistnonrel}
\end{equation}

To summarize, we have proven that the scattering by point defects
in bilayer graphene is more efficient than both in single-layer
graphene and in conventional electron gas. The difference with the
single-layer case is just due to vanishing density of states for
the massless Dirac fermions whereas for the bilayer graphene (as
well as for the conventional electron gas) it is constant.
However, for the two-dimensional nonrelativistic electrons an
arbitrary weak potential leads to formation of a bound state in
the gap~\cite{LL} which results in the logarithmic singularity of
the scattering amplitude at small energies (see
Eq.(\ref{resistnonrel})). In the case of the bilayer, there is no
gap and thus no localized states. As a result, the resistivity
should be just inversely proportional to the Fermi energy, or,
equivalently, to the charge carrier concentrations. This seems to
be in agreement with the recent experimental data \cite{wl1}.

I am thankful to Andre Geim and Kostya Novoselov for valuable discussions
stimulating this work. This work was supported by the Stichting voor
Fundamenteel Onderzoek der Materie (FOM) which is financially supported by
the Nederlandse Organisatie voor Wetenschappelijk Onderzoek (NWO).


\begin{thebibliography}{99}
\bibitem{misprint}  A misprint in Eq.(51) of Ref.\onlinecite{SSC} is fixed in Eq.(\ref{resistnonrel}).
\end{thebibliography}
\end{document}